# Online Monaural Speech Enhancement Based on Periodicity Analysis and A Priori SNR Estimation

Zhangli Chen* and Volker Hohmann

*Abstract*—This paper describes an online algorithm for enhancing monaural noisy speech. Firstly, a novel phase-corrected low-delay gammatone filterbank is derived for signal subband decomposition and resynthesis; the subband signals are then analyzed frame by frame. Secondly, a novel feature named periodicity degree (PD) is proposed to be used for detecting and estimating the fundamental period ($P_0$) in each frame and for estimating the signal-to-noise ratio (SNR) in each frame-subband signal unit. The PD is calculated in each unit as the multiplication of the normalized autocorrelation and the comb filter ratio, and shown to be robust in various low-SNR conditions. Thirdly, the noise energy level in each signal unit is estimated recursively based on the estimated SNR for units with high PD and based on the noisy signal energy level for units with low PD. Then the *a priori* SNR is estimated using a decision-directed approach with the estimated noise level. Finally, a revised Wiener gain is calculated, smoothed, and applied to each unit; the processed units are summed across subbands and frames to form the enhanced signal. The $P_0$ detection accuracy of the algorithm was evaluated on two corpora and showed comparable performance on one corpus and better performance on the other corpus when compared to a recently published pitch detection algorithm. The speech enhancement effect of the algorithm was evaluated on one corpus with two objective criteria and showed better performance in one highly non-stationary noise and comparable performance in two other noises when compared to a state-of-the-art statistical-model based algorithm.

*Index Terms*—Monaural speech enhancement, online implementation, periodicity analysis, *a priori* SNR estimation

## I. Introduction

ENHANCEMENT of speech from single-microphone recordings of speech in noisy environments is a challenging research topic. To solve this problem, many algorithms based on different frameworks have been developed. Among those algorithms reviewed in [1], the algorithms based on a statistical framework (also known as statistical-model based algorithms) perform consistently best on subjective speech quality evaluation across different noise conditions [2]. The statistical framework assume that the real

This work was supported by the DFG Cluster of Excellence EXC 1077/1 "Hearing4all". *Asterisk indicates corresponding author.*

The authors are with Medizinische Physik and Cluster of Excellence Hearing4all, University of Oldenburg, Oldenburg, 26129, Germany. E-mails: chenzl03@gmail.com, volker.hohmann@uni-oldenburg.de.

part and imaginary part of the fast Fourier transform (FFT) coefficient of the speech and the noise signals are zero-mean Gaussian or generalized Gamma distributed random variables; these two variables are independent from each other and the independence is kept across time frames and frequency bins. The performance of the algorithms based on this framework strongly relies on the accuracy of the estimation of spectral noise power. It is easy to estimate the noise spectrum level with voice activity detection (VAD) algorithms when the noise is stationary. However, it is difficult to do that when the noise becomes non-stationary, in particular if the noise envelope fluctuations have similar characteristics to those of the speech [1].

To deal with the problem of suppressing non-stationary noise, many strategies have been proposed. One popular strategy is to develop a better tracking algorithm for non-stationary noise based on the statistical framework described above. A review of the progress of approach can be found in chapter 9 of [1] and chapter 6 in [3]. The algorithm described in [4] is among the best of this class of algorithms. However, all of the algorithms based on the statistical framework have to assume that the spectrum levels of the noise change slowly frame by frame to make them distinguished from the spectrum levels of the speech which change fast. This means that these algorithms are not able to track highly non-stationary noise.

Another popular strategy is to detect the speech instead of the noise to separate the speech from the non-stationary noise. The most prominent feature used for the speech tracking is periodicity in voiced frames. Acoustic analysis shows that about 75% of the speech in spoken English are voiced and periodic [5]. The voiced phonemes often have larger energy than unvoiced phonemes and are more robust in various noisy conditions. Some frameworks have been set up to analysis and separate periodic speech from background noise. One example is the time-domain filtering framework [6, 7] which estimates the gain based on correlation calculation similar to that in the Wiener filtering framework. The algorithm described in [7] has shown to outperform two representative statistical-model based algorithms in perceptual evaluation of speech quality (PESQ) score [8] in relative low SNR conditions. However, this algorithm takes perfect pitch information from the clean signal, and assumes the order of the harmonic model of voiced speech is known. It is not known how much the performance will degrade when this algorithm is applied on a noisy signal directly and blindly. Another example is the computational auditory scene analysis (CASA) framework [9] which decomposes the signal into auditory filterbank and groups the subband by the periodicity (or pitch in the perceptual definition) of speech. The algorithm described in [9] showed good results in the separation of voiced speech from various background noise and outperformed a representative statistical-model based algorithm. However, this algorithm used the information of all frames of the signal, which made it unsuitable for online processing. Meanwhile, the algorithm derived and applied binary gain for the enhancement. The binary gain produces enhanced speech with low sound quality when compared to that produced by continuous (or soft) gain [10, 11].

Inspired by the successful use of speech periodicity information for monaural speech enhancement, an algorithm based on periodicity analysis is proposed here. Different from the algorithm in [7], the proposed algorithm is applied on the noisy signal blindly. Different from the algorithm in [9], the proposed algorithm is an online algorithm that only uses the information of current and previous frames for processing and makes it ready for realtime implementation in hearing devices; meanwhile, the proposed algorithm aims to derive and apply continuous gain to produce enhanced speech with high sound quality.

One important part of the proposed algorithm, which focuses on the periodicity analysis and SNR estimation for voiced speech, was previously presented in [12]. Here, an extended version of this part is developed and described with more details, including the improvement and the evaluation of the pitch detection and estimation approach using periodicity analysis. Another important part of the proposed algorithm focuses on the noise level estimation. A novel method of *a priori* SNR estimation is presented, which makes the algorithm applicable for both unvoiced parts and voiced parts of the speech. Last but not least, a novel implementation of the auditory gammatone filterbank for signal decomposition and resynthesis is introduced, which makes the algorithm suitable for online processing.

This paper is organized as follows. Section II introduces the details of the algorithm, mainly describing four parts: Signal decomposition and resynthesis, periodicity analysis, noise level and *a priori* SNR estimation, and gain calculation and application. Section III describes the evaluation of the algorithm. The accuracy of the fundamental period detection and the total speech enhancement effect of the proposed algorithm will be evaluated and compared with the performance of state-of-the-art reference algorithms. Section IV presents a discussion and the conclusions.

## II. Algorithm

Fig. 1 shows the block diagram of the proposed algorithm. Firstly, the signal is decomposed into frame-subband units. Secondly, the normalized autocorrelation (NAC) and the comb filtering ratio (CFR) are calculated and combined to form the periodicity feature, periodicity degree (PD), as a function of period candidates in each unit; the PD feature is analyzed across subbands in current and previous frames to detect and estimate the fundamental period ($P_0$) of the current frame; for the periodic frames (defined as the frames with detected $P_0$), the SNR of each unit is estimated based on PD and the estimated value of $P_0$ . Thirdly, the noise level of each unit is estimated from the estimated unit SNR in the periodic frames and by a recursive filtering of the noisy unit energy in the aperiodic frames (defined as the frames without detected $P_0$); from the estimated unit noise level in both periodic and aperiodic frames, the *a priori* SNR per unit is estimated. Finally, after applying the gain, the units are summed up across subbands and resynthesized across frames to form the enhanced signal; optionally, comb-filter post processing can be applied to further reduce the noise between the harmonics during the periodic frames.

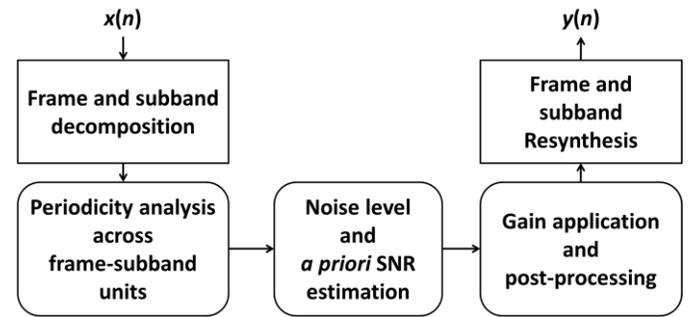

Fig. 1. Schematic diagram of the proposed algorithm.

### A. Signal Decomposition and Resynthesis

To simulate the human peripheral auditory filtering system, which we consider relevant to ensure close-to-optimum periodicity estimation, a gammatone filterbank [13] is used to decompose the signal into frequency subbands. However, the decomposed subband signal after gain application cannot be summed up directly to form the resynthesized signal because: 1) the peaks of the impulse responses of subband gammatone filters are not aligned; 2) the peak of the envelope of each subband impulse response is not aligned with the peak of its fine structure. To solve this problem, time-reversed filtering methods, e.g. [14], or phase-correction methods, e.g. [15], were applied in previous research.

In order to reduce the computational cost, the gammatone filter is often implemented in recursive form as infinite impulse response (IIR) filter. Holdsworth *et al.* [15] first introduced the digital approximation of the 4th-order gammatone filter by a cascade of 1st-order recursive filters. Hohmann [16] presented a more detailed implementation of the 4th-order gammatone filterbank. Different from previous implementations, Hohmann used the complex-valued expression of the gammatone filter, which brought two advantages: 1) the real part of the complex-valued filter output represents the desired subband signal, and the imaginary part represents the Hilbert transform of the desired subband signal; thus the absolute value of the complex-valued output represents the Hilbert envelope of the subband signal, which can be used for the following analysis and processing; 2) the alignment of the peaks of the envelope and the fine-structure of the impulse response can be easily achieved by multiplying the complex-valued subband signal by a fixed complex exponential factor.

Here a new implementation is derived. The advantages of using the complex-valued gammatone filter are adopted, and two improvements are introduced in the new implementation: 1) the numerator of the z-transform function of the gammatone filter is omitted in Hohmann's implementation [16] to make the implementation simpler, here the numerator will be kept to make the implementation more accurate; 2) the peak of the envelope of the gammatone impulse response is estimated from the Hilbert envelope of the subband signal in [16], here the peak is calculated directly by making the derivative of the expression of the envelope of the gammatone impulse response equal to zero. The details of the proposed implementation are explained below.

The z-transform of the 4th-order gammatone filter is [13, 16]:

$$G(k,z) = B(k) \cdot \frac{A(k)z^{-1} + 4A(k)^2 z^{-2} + A(k)^3 z^{-3}}{(1 - A(k)z^{-1})^4} \cdot C(k) \cdot z^{-D(k)} \quad (1)$$

where $k$ is the subband index, $A(k)$ is a complex-valued parameter decided by the center frequency (CF) and the equivalent rectangular bandwidth of subband $k$, $B(k)$ is the normalized gain, $C(k)$ is the phase shift for the fine structure to align the peak of fine structure and the peak of envelope, and $D(k)$ is the group delay of the whole filterbank to align the peaks of impulse response across subbands. The four parameters can be calculated by the following equations:

$$A(k) = exp\left\{-\frac{2\pi \cdot 1.019 ERB(k)}{f_s}\right\} \cdot exp\left\{i \cdot \frac{2\pi f_c(k)}{f_s}\right\} \quad (2)$$

$$B(k) = \sqrt{2} \cdot ERBstep \cdot \frac{(1 - A(k))^4}{A(k) + 4A(k)^2 + A(k)^3} \quad (3)$$

$$C(k) = exp\left\{-i \cdot \frac{2\pi f_c(k)}{f_s} \cdot min[N_{GD}, N_{PE}(k)]\right\} \quad (4)$$

$$D(k) = max[0, (N_{GD} - N_{PE}(k))] \quad (5)$$

In equation (2) and (4), $exp$ is exponential function, $i$ is the complex-valued unit, $f_s$ is sampling frequency, and $f_c(k)$ is the CF of subband $k$. In equation (2), the constant 1.019 [15] represents the ratio between the equivalent rectangular bandwidth of the gammatone filter and the equivalent rectangular bandwidth $ERB(k)$ of human auditory filters estimated from experimental data [17]. In equation (3), $ERBstep$ is the ratio between the frequency distance of adjacent CFs and $ERB(k)$, which should be equal to or smaller than 1 to ensure the filterbank covers the whole signal spectra. The relation between $ERB(k)$, $ERBstep$, and $f_c(k)$ is given by:

$$ERB(k) = 0.108 \cdot f_c(k) + 24.7 \quad (6)$$

$$ERBstep = \frac{f_c(k+1) - f_c(k)}{ERB(k)} \quad (7)$$

When the first CF, $f_c(1)$, the $ERBstep$ and the total number of filters are chosen, all other $f_c(k)$ can be calculated. The highest CF, $f_c(K)$, should be less than $f_s/2$.

In equations (4) and (5), $N_{GD}$ is the desired group delay of the filterbank. The choice of this parameter mainly affects the performance in low frequency subbands. Choosing $N_{GD}$ as a value corresponding to 16 milliseconds (ms) leads to a perfect resynthesis (as shown in Fig. 2). A smaller $N_{GD}$ (8 ms or 4 Hz) may be chosen to achieve lower processing delay; however, this may slightly distort the quality of the low frequency components of the resynthesized signal [16]. $N_{PE}(k)$ is the sample number corresponding to the peak position of the envelope of the impulse response. By taking the derivative of the envelope expression of the 4th-order gammatone filter equal to zero, $N_{PE}(k)$ can be calculated as:

$$N_{PE}(k) = round\left[\frac{3f_s}{2\pi \cdot 1.019 ERB(k)}\right], \quad (8)$$

where $round$ means rounding the value towards the nearest integer.

In summary, when the four parameters, including $f_c(1)$, $ERBstep$, filter number, and $N_{GD}$, are chosen, the coefficients of the complex-valued filterbank can be derived from the above equations. After applying the IIR filtering processing to the signal, the real part of the filtered complex-valued output is the subband signal, which can be summed up directly to form the resynthesized signal. The absolute value of the filtered complex-valued output forms the subband envelope which will be used in the following periodicity analysis in the subbands with CF larger than 1.5 kHz.

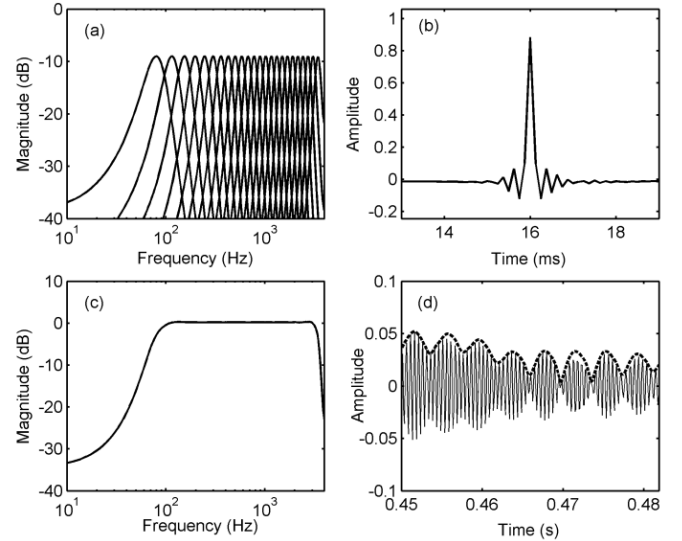

Fig. 2. Example of the proposed phase-corrected, complex-valued gammatone filterbank with following parameters: $f_c(1) = 80$ Hz, $ERBstep = 0.5$, filter number = 47, $N_{GD} = 128$. The sample frequency $f_s = 8$ kHz. (a) Frequency response of the filters; for clarity, only every second filter is displayed. (b) Overall impulse response of the analysis-resynthesis filterbank. (c) Frequency response of the overall impulse response. (d) The real part (thin solid line) and the absolute value (thick dashed line) of the complex-valued output of one filter with CF = 2032 Hz for a frame (32 ms) of the clean speech signal shown in Fig. 3.

Fig. 2 shows an example of the proposed implementation of a gammatone filterbank of 4th-order. The $f_s$ is 8 kHz, and the parameters are chosen as: $f_c(1) = 80$ Hz, $ERBstep = 0.5$, filter

number = 47, and $N_{GD} = 128$. So the highest CF $f_c(k = 47)$ is about 3440 Hz, which is smaller than $f_s/2$. Panel (a) shows the frequency response of each subband gammatone filter; for clarity, only every second filter is displayed. Panel (b) shows the overall impulse response of the analysis-resynthesis filterbank, which is calculated by summing the impulse responses of all subband filters. The peak at 16 ms is consistent with the chosen $N_{GD}$. Panel (c) shows the frequency response of the overall impulse response (b), which is perfectly flat across CFs. Panel (d) shows the real part (thin solid line) and the absolute value (thick dashed line) of the complex-valued filtered output of a frame (32 ms) of the clean speech signal shown in Fig. 3 at the 38th subband ($f_c(38) = 2032$ Hz). The absolute value of the complex-valued filtered output accurately describes the envelope of the subband signal. The fundamental period of this frame is about 4 ms, and the envelope accurately describes the fundamental period.

In the proposed algorithm, the subband filtering is conducted sample by sample, and the group delay is chosen as 16 ms. The filtered samples are then grouped into frames with length of 32 ms. The consecutive frames are overlapped with length of 16 ms. As a result of the subband filtering and short-time rectangular windowing, the input signal is decomposed into two-dimensional frame-subband units. After the analysis stage, the units are multiplied with a normalized Hamming window. Each windowed unit is multiplied by the gain estimated for that unit (see below) and all units are then summed across subbands and overlapped frames to resynthesize the enhanced signal. The frame by frame processing makes the algorithm suitable for online processing.

### B. Periodicity Analysis

The purpose of the proposed periodicity analysis is to calculate the periodicity feature PD in each frame-subband unit, detect the periodic frames, estimate the value of $P_0$ in each periodic frame, and estimate an initial SNR of the frame-subband unit in the periodic frames based on the calculated value of PD and the estimated value of $P_0$.

#### 1) Periodicity Feature Calculation

Two methods, NAC and CFR, are combined as periodicity feature PD. NAC and CFR are applied on the frame-subband filtered output at CFs lower than or equal to 1.5 kHz and on the envelope of the output at CFs higher than 1.5 kHz. The reason to analyze the envelope but not original waveform at high CF subbands is that the harmonics are usually unresolved in gammatone filters with CFs larger than 1.5 kHz. Some research has showed that the envelope which represents the amplitude modulation pattern of speech is more robust in $P_0$ estimation in noisy conditions than estimation from the resolved harmonic at lower frequencies [18].

Let $s(m)$ denote the clean speech signal, $d(m)$ denote the interference signal, and $x(m)$ denote the noisy speech signal. $d(m)$ is assumed to be an additive aperiodic noise and uncorrelated with $s(m)$:

$$x(m) = s(m) + d(m),\qquad(9)$$

where $m$ is the sample index of the whole signal.

For each frame-subband unit, NAC can be calculated as:

$$NAC(j,k,p) = \begin{cases} \dfrac{\sum_{n=0}^{N-1-p}[x(j,k,n)x(j,k,n+p)]}{\sqrt{\sum_{n=0}^{N-1-p}x(j,k,n)^2}\cdot\sqrt{\sum_{n=0}^{N-1-p}x(j,k,n+p)^2}}, & k\in K_L \\[3ex] \dfrac{\sum_{n=0}^{N-1-p}[x_E(j,k,n)x_E(j,k,n+p)]}{\sqrt{\sum_{n=0}^{N-1-p}x_E(j,k,n)^2}\cdot\sqrt{\sum_{n=0}^{N-1-p}x_E(j,k,n+p)^2}}, & k\in K_H \end{cases}$$

(10)

where $j$ and $k$ are the frame and subband indexes, $K_L$ is the set of subband indexes for CFs lower than or equal to 1.5 kHz, $K_H$ is the set of subband indexes for CFs higher than 1.5 kHz, $p$ is the period candidate (in samples), $n$ is the sample index of the frame signal, $N$ is the frame length, $x_E$ is the signal envelope which has been normalized to zero mean. The fundamental frequency ($F_0$) is searched in the range between 70 Hz and 420 Hz in the proposed algorithm. So the $P_0$ is searched in the range from 2.4 ms to 14.3 ms. When $f_s$ is 8 kHz, $p$ is in the range of 19 to 114.

A simple method, the average magnitude difference function (AMDF) [19], has been found effective in the $P_0$ detection and estimation for clean speech. The AMDF is the absolute magnitude of the difference between the original signal and its delayed version, and exhibits a notch at the delay corresponding to $P_0$. Here, a variation of AMDF, CFR, is defined as the ratio of the frame energy of the summation between the original signal and its delayed version to the frame energy of the difference between the original signal and its delayed version, and calculated as:

$$CFR(j,k,p) = \begin{cases} \dfrac{\sum_{n=0}^{N-1-p}[x(j,k,n)+x(j,k,n+p)]^2}{\sum_{n=0}^{N-1-p}[x(j,k,n)-x(j,k,n+p)]^2}, & k\in K_L \\[3ex] \dfrac{\sum_{n=0}^{N-1-p}[x_E(j,k,n)+x_E(j,k,n+p)]^2}{\sum_{n=0}^{N-1-p}[x_E(j,k,n)-x_E(j,k,n+p)]^2}, & k\in K_H \end{cases}$$

(11)

Differently from AMDF, CFR will exhibit a peak at the delay corresponding to $P_0$.

Previous research [20] showed that combining the methods of autocorrelation and AMDF can improve the accuracy of pitch detection and estimation for noisy speech. Recently Tan and Alwan [21] proposed a multi-band summary correlogram (MBSC) based pitch detection algorithm for noisy speech. The MBSC algorithm calculated the harmonic-to-subharmonic energy ratio (HSR) by comb filter in frequency domain and used this ratio to weight the autocorrelation to achieve a peak-enhanced summary correlogram and improve the pitch detection. The HSR is similar to the CFR described here. According to the successful approaches mentioned above, NAC and CFR are combined here as the periodicity feature PD:

$$PD(j,k,p) = max[0.01, \ NAC(j,k,p)\cdot CFR(j,k,p)], \qquad(12)$$

where $max$ means taking the maximum value of the two values in the bracket.

Fig. 3 shows an example of the calculations of periodicity features including NAC, CFR, and PD. The example sentence and noise are from the NOIZEUS corpus [1]. The clean speech is a female-spoken sentence (named sp24 in the corpus: *The drip of the rain made a pleasant sound*), the noise is highly non-stationary train noise, and the overall SNR of the noisy signal is 0 dB. Panel (a) and (b) show the spectrograms of the clean signal and the noise signal, respectively; the color-bar on the right site of the spectrogram is in dB scale. Panel (c) shows the subband-averaged NAC of the noisy signal, which is calculated by averaging $NAC(j, k, p)$ in (10) across subbands. Panel (d) and (e) show the subband-averaged CFR and subband-averaged PD, respectively, which are calculated in the same way as the subband-averaged NAC. Compared to (c), (d) shows a better resolution of the periodicity feature (e.g., at the time around 1.3 s); however, (d) also shows more subharmonic (e.g., at the time around 1.8 s) which may increase the difficulty of $P_0$ detection. Panel (e) shows the result of the combination of (c) and (d).

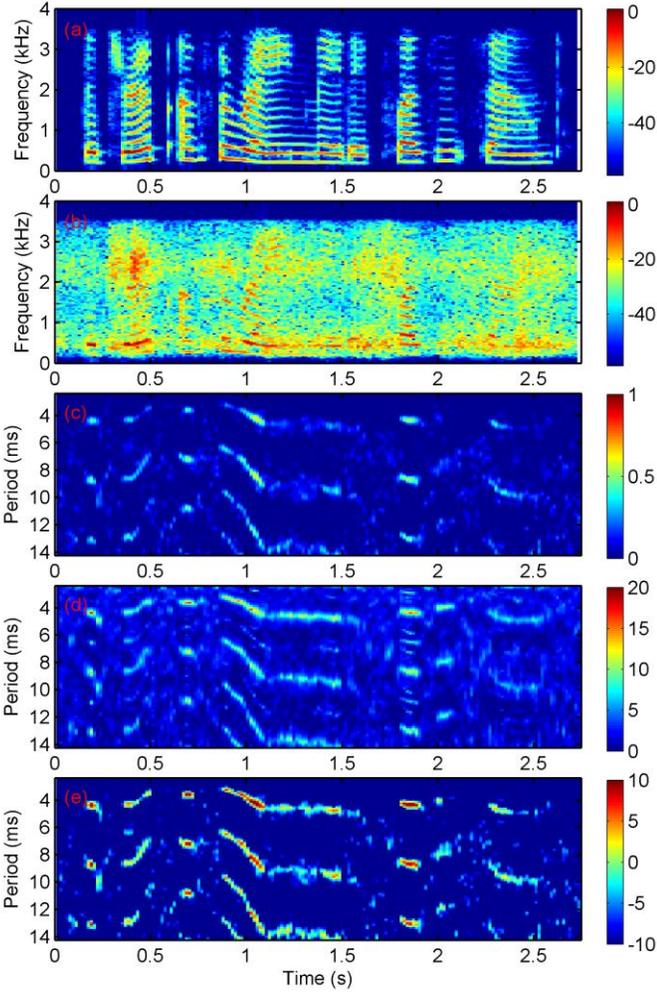

Fig. 3. Example of periodicity feature calculation. (a) Spectrogram of the clean signal. (b) Spectrogram of the noisy signal (in train noise with overall SNR = 0 dB). (c) Subband-averaged NAC of the noisy signal. (d) Subband-averaged CFR of the noisy signal. (e) Subband-averaged PD of the noisy signal.

### 2) $P_0$ Detection and Estimation

The subband-averaged PD described above is used to detect the periodic frames and estimate the value of $P_0$. However, some random blocks (e.g., at the time around 0.1 s or 2.7 s) still exist, which mainly stem from dominant stationary parts of the noise and may trigger false alarms in the $P_0$ detection. Here a simple method is applied to reduce the contribution of the stationary part of the noise to the subband-averaged PD, which is found to be effective to suppress these random blocks.

Firstly, a simple onset detection method is applied to estimate the energy level (which is calculated as the sum of the absolute squares of signal samples in each frame-subband unit) of the stationary part of the noise. For the first frame, the energy level of noise in each frame-subband unit is assumed to be equal to the energy level of the noisy signal; from the second frame on, the following iteration is applied:

$$\hat{E}_{0d}(j, k) = \begin{cases} \hat{E}_{0d}(j-1, k), & when \left[\dfrac{E_x(j, k)}{\hat{E}_{0d}(j-1, k)}\right] > \delta \\ \alpha \hat{E}_{0d}(j-1, k) + (1-\alpha)E_x(j, k), & when \left[\dfrac{E_x(j, k)}{\hat{E}_{0d}(j-1, k)}\right] \leq \delta \end{cases}$$

(13)

where $E_x$ is the energy level of the noisy signal, and $\hat{E}_{0d}$ is the estimated energy level of the stationary part of the noise; the recursive smooth parameter $\alpha$ and the threshold parameter $\delta$ are empirically chosen as 0.96 and 1.4, respectively.

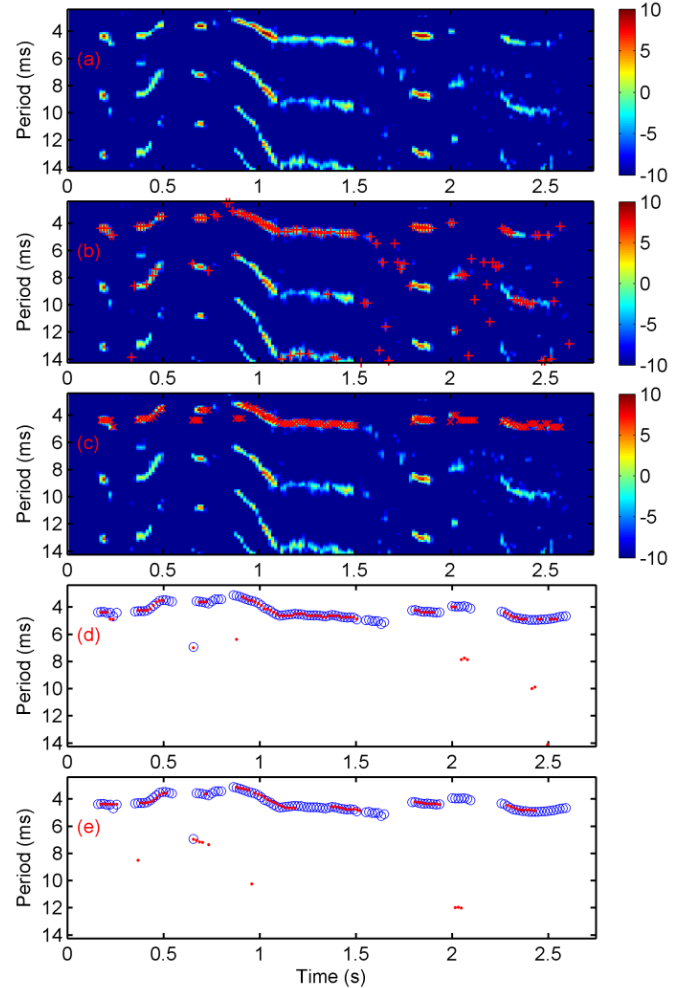

Fig. 4. $P_0$ detection for the noisy speech in Fig. 3. (a) EFPD (Subband-averaged of PD with $CG_1$ weighting). (b) Maximum peaks above the preset threshold (as the "+" labels) of EFPD. (c) The detected memory-$P_0$ (as the "x" labels). (d) $P_0$ detection result by the proposed method. (e) $P_0$ detection result by a recently published method [21]. The circles denote the ground truth.

Based on the above estimation of stationary noise level, an initial SNR based on maximum likelihood estimation can be calculated as:

$$\widehat{SNR}_0(j,k) = \frac{E_x(j,k)}{\hat{E}_{0d}(j,k)} - 1, \tag{14}$$

and an initial Wiener gain [1] can be calculated as:

$$G_0(j,k) = \frac{\widehat{SNR}_0(j,k)}{\widehat{SNR}_0(j,k) + 1}. \tag{15}$$

The PD of each unit is then weighted with the initial Wiener gain and averaged across subbands to form the enhanced frame periodicity degree (EFPD):

$$EFPD(j,p) = \frac{1}{K}\sum_{k=1}^{K}[G_0(j,k) \cdot PD(j,k,p)] \tag{16}$$

Panel (a) of Fig. 4 shows the EFPD of the noisy signal in Fig. 3. Compared to the PD in panel (e) of Fig. 3, it can be seen that most of the random blocks have been suppressed in EFPD.

To detect the periodic frames and estimate the value of $P_0$, an intuitive method is to detect the maximum value in each frame of the EFPD: if this maximum value is above a preset PD threshold, this frame is detected as a potential periodic frame, and the estimated value of $P_0$ is equal to the period candidate corresponding to this maximum value. However, this simple method may produce many subharmonic or harmonic errors. As shown in panel (b) of Fig. 4, some maximum values above the preset threshold (plotted as the "+" labels) appear at the second (e.g., at the time around 0.45 s or 2.4 s) or the third (e.g., at the time around 1.2 s) subharmonic. To reduce these errors, an online tracking algorithm, which only used the information of current and previous frames, is applied here. The online tracking algorithm consists of four main steps: adaptive dual PD thresholding to detect the potential periodic frames, EFPD peak detection to locate the period candidates of $P_0$, memory-$P_0$ estimation to restrict the search range of $P_0$ and reduce the harmonic and subharmonic errors, and continuous tracking to ultimately decide the frame as periodic or aperiodic and estimate the value of $P_0$. The details of the four steps are described below.

In the first step, before calculating the adaptive dual PD threshold, the adaptive dual SNR threshold is calculated based on the subband average of the initial SNR in (14):

$$SNRTHD_1(j)$$
$$= max\left[0.3, \left\{0.6 + 0.03\left(min\left[30, max\left[0, 10 \cdot \lg\left(\overline{FSNR}_0(j)\right)\right]\right] - 10\right)\right\}\right] \tag{17}$$

$$SNRTHD_2(j)$$
$$= max\left[0.1, \left\{0.2 + 0.01\left(min\left[30, max\left[0, 10 \cdot \lg\left(\overline{FSNR}_0(j)\right)\right]\right] - 10\right)\right\}\right] \tag{18}$$

where $\overline{FSNR}_0$ is the subband average of $\widehat{SNR}_0$ in (14). $SNRTHD_1$ is the upper threshold and in the range between 0.3 (when $10\lg\left(\overline{FSNR}_0\right) \leq 0\ dB$) and 0.9 (when $10\lg\left(\overline{FSNR}_0\right) \geq 30\ dB$). $SNRTHD_2$ is the lower threshold and in the range between 0.1 (when $10\lg\left(\overline{FSNR}_0\right) \leq 0\ dB$) and 0.2 (when $10\lg\left(\overline{FSNR}_0\right) \geq 30\ dB$). Then the adaptive dual PD threshold is calculated according to the relationship between SNR and PD as described in (24). When $SNRTHD_1$ is in the range between 0.3 and 0.9, the PD threshold, $PDTHD_1$, is in the range between 0.37 (−4.3 dB) and 1.3 (1.2 dB); when $SNRTHD_2$ is in the range between 0.1 and 0.2, $PDTHD_2$ is in the range between 0.11 (−9.6 dB) and 0.23 (−6.3 dB). The constants in (17) and (18) are chosen empirically.

In the second step, the local peaks in EFPD are detected. The frames with peaks larger than $PDTHD_2$ are defined as potential periodic frames. Only these potential periodic frames are used in the following two steps for the detection of periodic frames and the estimation of $P_0$.

In the third step, the memory-$P_0$ of each potential periodic frame is estimated as the median value of the period corresponding to the maximum peak of 50 previous frames whose maximum peak is larger than $PDTHD_1$. If the deviation of the period corresponding to the maximum peak in the current frame and the memory-$P_0$ is smaller than 40% of the memory-$P_0$, the memory-$P_0$ is further updated to the period corresponding to the maximum peak in the current frame. An example of memory-$P_0$ detection is shown in panel (c) of Fig. 4 (as the "x" labels). It can be seen that the estimated memory-$P_0$ in each frame is well consistent with the true $P_0$ (as the circles, which are obtained by analyzing the clean speech with the software Praat [22] and with some additional manual correction).

In the final step, the $P_0$ is detected and estimated in each potential periodic frame according to its continuity property: if the previous potential periodic frame does not have a detected $P_0$, for the current potential periodic frame, a $P_0$ is detected only when the maximum peak is above $PDTHD_1$, and the estimated value of $P_0$ equals to the period corresponding to the maximum peak; if the previous potential periodic frame has a detected $P_0$, for the current potential periodic frame, a $P_0$ is detected, and the estimated value of $P_0$ equals to the period of the peak which is closest to memory-$P_0$. The continuous tracking result (as the dots) is shown in panel (d) of Fig. 4. The true $P_0$ of each frame (the circles) is also shown in the figure. For comparison, the detected result (as the dots) of a recently published multi-band summary correlogram-based (MBSC) pitch detection algorithm [21] is also shown in panel (e). The $P_0$ detection error rate (defined at part A of section III) is 23.4% for the proposed algorithm and 32.2% for the MBSC algorithm for this example. A more comprehensive comparison of the two algorithms is described in part A of section III.

### 3) Subband SNR Estimation of Periodic Speech Frames

For the periodic frames, the SNR of each frame-subband unit can be estimated from the PD and the estimated value of $P_0$, by assuming that: 1) the speech is uncorrelated with the interference; 2) the interference is uncorrelated with its delayed version. Below shows how to derive the relationship between PD, estimated $P_0$, and SNR based on the above two assumptions.

When the period candidate $p$ equals to true $P_0$, for the subbands with CF lower than 1.5 kHz, the NAC in (10) can be expressed approximately as [23]:

$$NAC(j,k,P_0(j)) \approx \frac{\sum_{n=0}^{N-1} s(j,k,n)^2}{\sum_{n=0}^{N-1} s(j,k,n)^2 + \sum_{n=0}^{N-1} d(j,k,n)^2}, \quad k \in K_L \tag{19}$$

where $P_0(j)$ is the true $P_0$ at frame $j$. For the Hilbert envelope of the signal, the above two uncorrelated assumptions are approximately kept; meanwhile, the energy level of the Hilbert envelope is two times of the energy level of the original signal, so for the subbands with CF higher than 1.5 kHz:

$$NAC(j,k,P_0(j)) \approx \frac{\sum_{n=0}^{N-1} s_E(j,k,n)^2}{\sum_{n=0}^{N-1} s_E(j,k,n)^2 + \sum_{n=0}^{N-1} d_E(j,k,n)^2}$$
$$\approx \frac{\sum_{n=0}^{N-1} s(j,k,n)^2}{\sum_{n=0}^{N-1} s(j,k,n)^2 + \sum_{n=0}^{N-1} d(j,k,n)^2}, \quad k \in K_H \tag{20}$$

where $s_E$ and $d_E$ denote the Hilbert envelope of signal $s$ and $d$, respectively. The SNR of each frame-subband unit is defined as:

$$SNR(j,k) = \frac{\sum_{n=0}^{N-1} s(j,k,n)^2}{\sum_{n=0}^{N-1} d(j,k,n)^2} \tag{21}$$

So (19) and (20) can be combined and expressed as:

$$NAC(j,k,P_0(j)) \approx \frac{SNR(j,k)}{SNR(j,k)+1} \tag{22}$$

Similarly, when the period candidate $p$ equals the true $P_0$, the CFR in (11) can be expressed as:

$$CFR(j,k,P_0(j)) \approx 2SNR(j,k)+1 \tag{23}$$

and PD in (12) can be expressed as:

$$PD(j,k,P_0(j)) \approx max\left[0.01, \left(\left[\frac{SNR(j,k)}{SNR(j,k)+1}\right] \cdot [2SNR(j,k)+1]\right)\right] \tag{24}$$

By replacing the true $P_0$, $P_0(j)$, with the estimated $P_0$, $\hat{P}_0(j)$, the $PD\left(j,k,\hat{P}_0(j)\right)$ can be calculated by (12). By solving (24), the SNR of each frame-subband unit in the periodic frames can be estimated as:

$$\widehat{SNR}_v(j,k)$$
$$\approx \frac{1}{4}\left[PD\left(j,k,\hat{P}_0(j)\right)-1+\sqrt{PD\left(j,k,\hat{P}_0(j)\right)^2+6PD\left(j,k,\hat{P}_0(j)\right)+1}\right] \tag{25}$$

For ideal conditions, like voiced speech in white noise, the above two uncorrelated assumptions are well satisfied. One example can be found in [12]. For other non-ideal conditions, like speech in multi-speaker interference, the two assumptions are not fully satisfied and the accuracy of SNR estimation will be degraded.

Fig. 5 shows the results of SNR estimation of the frames detected as periodic of the noisy speech in Fig. 3. In each panel, the line shows the theoretical relation and the dots show the calculated relation between the periodicity features (NAC, CFR, and PD) and the true (known) SNR in each frame-subband unit. The x-axis values of the line or the dots are the true SNR (when the clean signal and the noise are known) in each frame-subband unit. In panel (a), (b), and (c), the y-axis values of the lines are $NAC(j,k,P_0(j))$ calculated by (22), $CFR(j,k,P_0(j))$ calculated by (23), and $PD(j,k,P_0(j))$ calculated by (24), respectively; the y-axis values of the dots are $NAC(j,k,p)$ calculated by (10) when $p = \hat{P}_0(j)$, $CFR(j,k,p)$ calculated by (11) when $p = \hat{P}_0(j)$, and $PD(j,k,p)$ calculated by (12) when $p = \hat{P}_0(j)$, respectively. It can be seen that the dots are scattering around the lines, which shows the accuracy of the SNR estimation from the respective features on the level of frame-subband units. In panel (d), the line is a straight line (which means the estimated SNR should equal to the true SNR in theoretical) and the y-axis values of the dots are the estimated SNR calculated by (25). It can be seen that the estimated SNR has a good linear relation with true SNR, especially for the unit whose true SNR is larger than 0 dB. For the some unit whose true SNR is smaller than 0 dB, the estimated SNR is larger than the true SNR but kept below 0 dB.

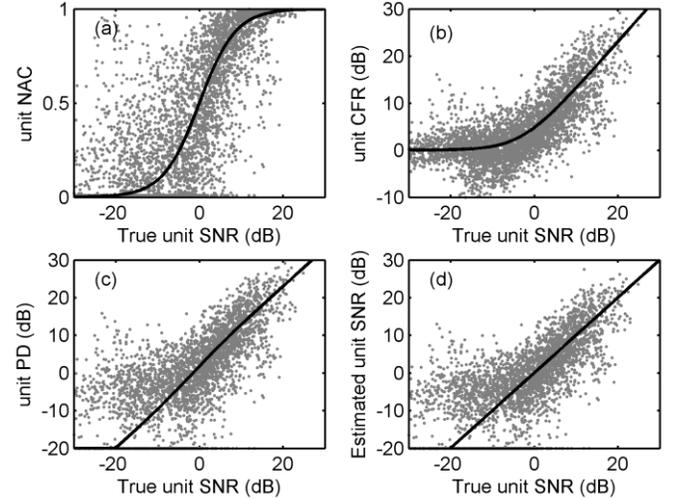

Fig. 5. SNR estimation of the frames detected as periodic of the noisy speech in Fig. 3. In all panels, the x-axis values of the dots are the true SNR in each frame-subband unit. (a) The line shows the theoretical relationship between NAC and true SNR, as calculated by (22); the y-axis values of the dots are $NAC(j,k,p)$ calculated by (10) when $p = \hat{P}_0(j)$. (b) The line shows the theoretical relationship between CFR and true SNR, as calculated by (23); the y-axis values of the dots are $CFR(j,k,p)$ calculated by (11) when $p = \hat{P}_0(j)$. (c) The line shows the theoretical relationship between PD and true SNR, as calculated by (24); the y-axis values of dots are $PD(j,k,p)$ calculated by (12) when $p = \hat{P}_0(j)$. (d) The line is a straight line; the y-axis values of the dots are the estimated SNR calculated by (25).

### C. Estimation of Noise Level and a priori SNR

The periodicity analysis stage can estimate the SNR of the frame-subband units in the periodic frames of the noisy speech, but cannot deal with units in the aperiodic frames of the noisy speech. To deal with both the periodic and aperiodic frames of the speech, a processing stage similar to the classical methods of noise level estimation (e.g., chapter 9 in [1]) and a priori SNR estimation [24] is proposed here.

For the aperiodic frames, the noise energy level of each frame-subband unit is estimated by a recursive filter:

$$\hat{E}_d(j,k) = min\left[E_x(j,k), \ \beta_1\hat{E}_d(j-1,k) + (1-\beta_1)E_x(j,k)\right] \tag{26}$$

where $\beta_1$ is a smoothing factor and empirically selected as 0.9. Then the speech energy level is estimated by the decision-directed approach [24] as:

$$\hat{E}_s(j,k) = min\{E_x(j,k),\ \beta_2[G(j-1,k)\cdot E_x(j-1,k)] + (1-\beta_2)[E_x(j,k)-\hat{E}_d(j,k)]\} \tag{27}$$

where $\beta_2$ is a smoothing factor and empirically selected as 0.96, and $G$ is the final Wiener gain calculated as in (30). On the right side of (27), the first term $[G(j-1,k)\cdot E_x(j-1,k)]$ can be seen as the estimated speech energy level in the previous frame without smoothing; the second term $[E_x(j,k)-\hat{E}_d(j,k)]$ can be seen as the maximum likelihood estimation of the speech energy level in current frame, which will always be larger than zero as $\hat{E}_d$ has been limited to $E_x$ in (26).

For the periodic frames, the noise level is estimated in two ways according to the estimated SNR calculated in (25). For the units with $\widehat{SNR}_v(j,k)$ larger than or equal to 1, which means that these units show obvious periodicity, the noise energy level is estimated from $\widehat{SNR}_v(j,k)$ and the noisy energy $E_x(j,k)$:

$$\hat{E}_d(j,k) = min\left\{E_x(j,k), \beta_3\hat{E}_d(j-1,k) + (1-\beta_3)\left[\frac{E_x(j,k)}{\widehat{SNR}_v(j,k)+1}\right]\right\} \tag{28}$$

where $\beta_3$ is empirically selected as 0.9. For the units with $\widehat{SNR}_v(j,k)$ smaller than 1, which means these units are dominated by the aperiodic noise or the aperiodic components of the imperfect voiced speech, an initial noise energy level is estimated by (26) firstly. Then this initial noise energy is compared with the noisy energy $E_x(j,k)$; if $E_x(j,k)$ is less than two times of the initial noise energy, the estimated noise energy equals to the initial noise energy; if $E_x(j,k)$ is less than two times of the initial noise energy, the estimated noise energy is calculated by (26) but with $\beta_1$ empirically selected as 0.8. A smaller value of $\beta_1$ performs a faster tracking than a larger value. Then the speech energy level is calculated by (27) but with $\beta_2$ empirically selected as 0.8. Again, a smaller value of $\beta_2$ performs a faster tracking than a larger value. The better tracking effect of speech component with lower smoothing parameters has also been shown in [25].

For all frames, the *a priori* SNR is estimated as:

$$\widehat{SNR}(j,k) = \frac{\hat{E}_s(j,k)}{\hat{E}_d(j,k)} \tag{29}$$

Previous research shows that directed-decision approach to estimate the *a priori* SNR is key to reduce the musical noise [26]. An informal listening test of the proposed algorithm confirmed the suppression of musical noise by this approach.

Fig. 6 shows noise level estimation result of the noisy signal in Fig. 3. Panel (a) shows the cochleagram, i.e., the sub-band amplitude of the gammatone filterbank output as a function of time and frequency of the true noise; the color bar on the right side is in dB scale. It can be seen that the noise is highly nonstationary. Panel (b) shows the noise cochleagram estimated by the proposed algorithm. To have a better view of the comparison, the true and estimated noise levels in a low-CF subband and a high-CF subband are shown in panels (c) and (d), respectively. The dashed lines represent the true noise levels and the solid lines represent the estimated noise levels. It can be seen that the estimated noise levels well follow the sudden changes of the true noise levels. Please note that here the estimated noise levels are compared with the true noise levels, not with the recursive filtering smoothed levels of the true noise as in [4].

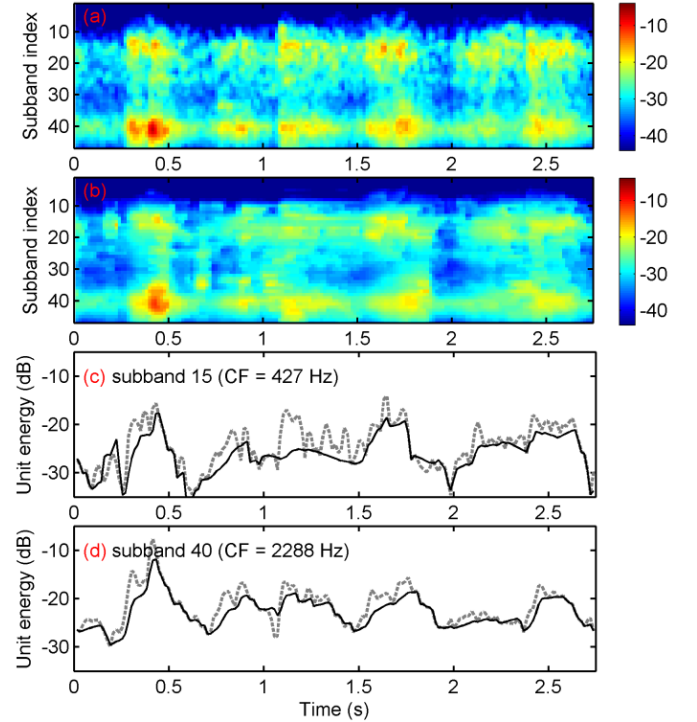

Fig. 6. Noise-level estimation of the noisy signal in Fig. 3. (a) The cochleagram of the highly nonstationary train noise. (b) The cochleagram of the noise estimated by (26) and (28). (c) The true (dashed line) and the estimated (solid line) noise level in a low-CF subband (index = 15, CF = 427 Hz). (d) The true (dashed line) and the estimated (solid line) noise level in a high-CF subband (index = 40, CF = 2288 Hz).

### D. Gain Calculation

With the SNR of each frame-subband unit estimated in previous stage, a continuous gain is calculated as:

$$G(j,k) = max\left\{G_{min}, \left(\frac{\widehat{SNR}^2(j,k)}{\widehat{SNR}^2(j,k)+1}\right)\right\}, \tag{30}$$

where $G_{min}$ is the preset minimum gain and chosen as 0.178. As the gain will be applied to the subband signal directly, the dB scale of 0.178 is calculated as $20\cdot log10(0.178)$ and equal to -15 dB. The gain calculated by (30) is a revised form of the classical Wiener gain. This gain has a steeper transition compared to the Wiener gain and a smoother transition compared to binary masking gain. It is found that using this gain results in a better SNR improvement compared to using Wiener gain and meanwhile a better PESQ score compared to using binary masking gain.

Panel (b) in Fig. 7 shows the gain estimated by the proposed method for the noisy signal in Fig. 3. For comparison, the ideal (when true SNR is known) Wiener gain for the same signal is shown in panel (a). It can be seen that the estimated gain resembles the ideal gain well for the voiced frames. The differences between the estimated gain and the ideal gain mainly occurs at the aperiodic frames (e.g., at the time around 0.3 s) and the very high CF subbands of the periodic frames (e.g., at the time around 1.8 s or 2.3 s).

There are some small random gain blocks at the aperiodic frames in panel (b). These random gain blocks may cause musical noise. To suppress these random blocks, a simple online smoothing method is applied here: for each aperiodic frame, if the gain of its previous frame $G(j-1,k)$ is smaller than 0.1, the gain of its next subbands, $G(j, k-1)$ and $G(j, k+1)$, are smaller than 0.3, and the gain of current frame $G(j,k)$ is smaller than 0.6, then $G(j,k)$ will be set to minimum gain. The estimated gain after smoothing is shown in panel (c). It can be seen that some small random gain blocks have been eliminated (e.g., at 0.6 s and 2.7 s).

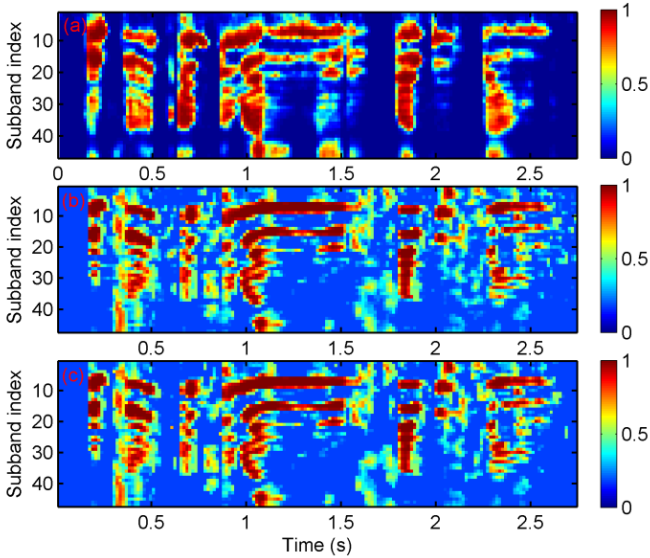

Fig. 7. Estimated gain for the noisy signal in Fig. 3. The color bar shows the value of the gain in all panels. (a) The ideal (when true SNR is known) Wiener gain. (b) The gain estimated by the proposed algorithm. (c) The gain estimated by the proposed algorithm after smoothing.

As the gammatone filters have spectral overlap between the adjacent subbands, the reconstructed signal may still contain some noise between the adjacent harmonics in periodic frames after applying the gain. Applying a simple feed-forward comb filter during the periodic frames may reduce this noise. For the periodic frames, the enhanced signal is further filtered as:

$$y(j,k,n) = \begin{cases} 0.5\left(x_G(j,k,n) + x_G\left(j,k,n+\hat{P}_0(j)\right)\right), & when\ n \leq \frac{N}{2} \\ 0.5\left(x_G(j,k,n) + x_G\left(j,k,n-\hat{P}_0(j)\right)\right), & when\ n > \frac{N}{2} \end{cases}$$ (31)

where $x_G(j,k,n)$ is the enhanced signal unit after applied the gain in (30). Please note that the maximal value of $\hat{P}_0(j)$ (corresponding to 14.3 ms) is smaller than $N/2$ (corresponding to 16 ms); and this comb-filtering will not introduce signal delay in this frame-based processing.

## III. EVALUATION

The performance of the proposed algorithm will be evaluated in two aspects: the accuracy of the $P_0$ detection and the objective scores of the speech enhancement effect.

The corpus used in both evaluations is the NOIZEUS corpus produced by Loizou [1]. This corpus contains 30 sentences spoken by three male and three female speakers and eight types of daily noise and has been used for subjective and objective evaluations of many speech enhancement algorithms [1]. Only three representative types of noise (car, train, and babble noise) will be used here. The car noise is relatively stationary and the train noise is highly non-stationary; the car noise and train noise are aperiodic and the babble noise contains periodic components.

### A. $P_0$ Detection Accuracy

The accuracy of the $P_0$ detection is essential for the total performance of the proposed algorithm. Before evaluating the whole enhancement effect of the proposed algorithm, the $P_0$ detection part is evaluated in comparison to a recently published multi-band summary correlogram-based (MBSC) algorithm [21]. The MBSC algorithm was compared with several representative algorithms in [21] and was shown to perform best. Therefore, it is used as a benchmark here. The implementation of the algorithm is a Matlab function "mbsc.m" that was downloaded from the official website of the authors.

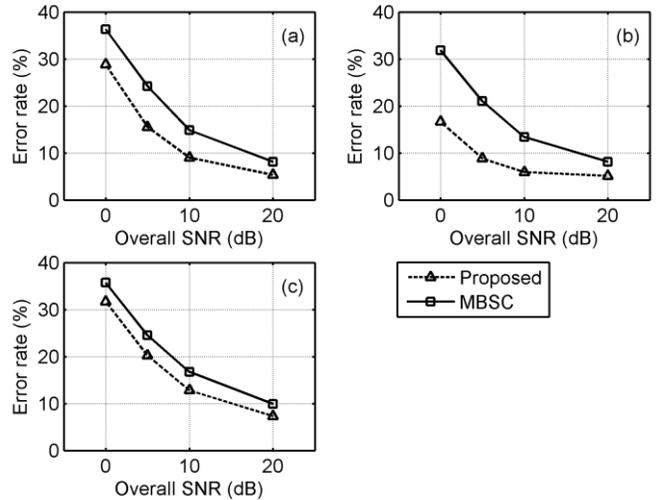

Fig. 8. Error rates of $P_0$ detection by the proposed algorithm (triangles) and the MBSC algorithm (squares) on the NOIZEUS corpus. (a) Speech in car noise. (b) Speech in train noise. (c) Speech in babble noise.

The sentences from the NOIZEUS corpus were mixed with the car, train, and babble noise at an overall SNR of 0, 5, 10, and 20 dB, respectively. The noisy signals were filtered by the modified Intermediate Reference Systems (IRS) filters used in ITU-T P.862 [8] to simulate the receiving frequency characteristics of telephone handsets. As the IRS filter has a flat bandpass response between 300 and 3400 Hz, the fundamental harmonics below 300 Hz of the speech are attenuated and this makes $P_0$ detection an even more challenging task. The reference $P_0$ was obtained by analyzing the clean sentences with the software Praat [22], with some additional manual correction.

Fig. 8 shows the error rates of $P_0$ detection by the proposed algorithm (triangles) and the MBSC algorithm (squares) on NOIZEUS corpus. The error rate is calculated as the percentage of misses (when a periodic frame is detected as aperiodic), false alarms (when an aperiodic frame is detected as periodic), and deviations (when the difference between the detected $F_0$ and the true $F_0$ is larger than 20% of the true $F_0$). This calculation method is the same as that in [21]. Panels (a), (b), and (c) show the results in car, train, and babble noise, respectively. It can be seen that the proposed algorithm outperforms the MBSC algorithm in all three noisy conditions.

The NOIZEUS corpus only has 30 sentences (the total time length is about 100 s) which may not be able to fully reveal the performance of the two algorithms. To further verify the accuracy of the $P_0$ detection part of the proposed algorithm, the Keele corpus [27] was also evaluated here. The Keele corpus contains a phonetically balanced story (about 30 s long) read by five female and five male speakers. This corpus is widely used in the evaluation of pitch detection algorithms. The sentences are down-sampled to 8 kHz and mixed with three real-world noise types – babble, car (Volvo), and machine gun at the overall SNR of 0, 10, and 20 dB. These noise files are from the NOISEX-92 corpus [28].

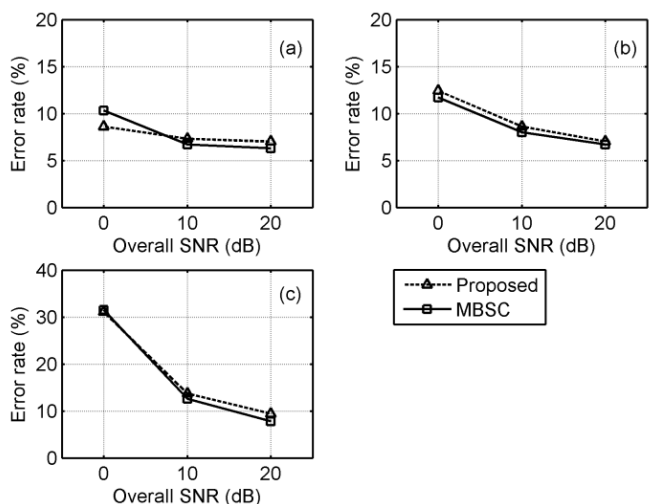

Fig. 9. Error rates of $P_0$ detection by the proposed algorithm (triangles) and the MBSC algorithm (squares) on Keele corpus. (a) Speech in Volvo car noise. (b) Speech in machinegun noise. (c) Speech in babble noise.

Fig. 9 shows the error rates of $P_0$ detection by the proposed algorithm (triangles) and the MBSC algorithm (squares) on Keele corpus. The results for the MBSC algorithm is very close to the results in [21], which validates the correct implementation of the MBSC algorithm here. It can be seen that the proposed algorithm has comparable performance as the MBSC algorithm on this corpus.

From the above two figures it can be seen that the MBSC algorithm performs well on full-band signals but poorly on the bandpass signals. However, the proposed algorithm performs well on both types of signals. Meanwhile, the frame length used in MBSC algorithm is adaptive between 10 ms and 80 ms. The longer frame length used in pitch detection may bring advantages to the MBSC algorithm in the evaluation. However, the longer frame length is not suitable for online processing. Thus, it is a positive result that the proposed algorithm with a short frame length (32 ms) achieves similar results as the MBSC algorithm for full-band signals and better results for bandpass signals.

## B. Speech Enhancement Effect

The speech enhancement effect of the proposed algorithm was mainly evaluated with two objective criteria which are usually used in the evaluation of speech enhancement algorithm: the overall SNR and the perceptual evaluation of speech quality (PESQ) score. The overall SNR can show the similarity between the enhanced signal and the clean signal. It was calculated as:

$$ovlSNR = 10\lg\left(\frac{\sum_n s^2(m)}{\sum_n[s(m) - y(m)]^2}\right), \quad (32)$$

where $y(m)$ is the enhanced signal. The PESQ has a higher correlation with speech quality than SNR [29]. Here, PESQ was calculated by the MATLAB function from the CD in [1].

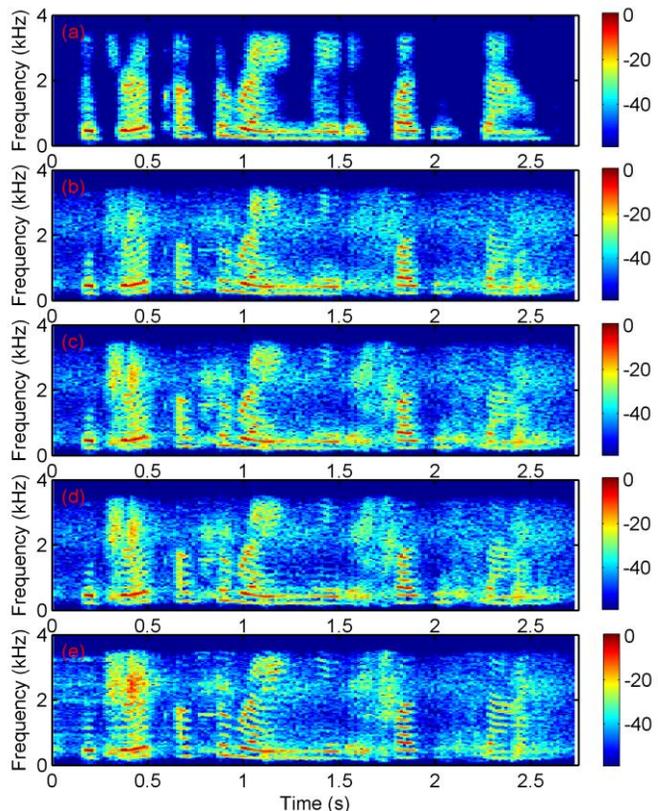

Fig. 10. Spectrogram of the enhancement result of the noisy signal in Fig. 3. (a) Result by ideal Wiener gain. (b) Result by the gain estimated from $\widehat{SNR}_v$ in (25). (c) Result by the gain calculated by (30). (d) Result by gain calculated by (30) plus comb-filtering by (31). (e) Result by the MMSE algorithm.

The proposed algorithm was evaluated with the NOIZEUS

corpus described above. One state-of-the-art statistical-model based minimum mean-square error (MMSE) monaural speech enhancement algorithm was also evaluated as a comparison. This MMSE algorithm includes a recently developed MMSE-based estimation algorithm of noise power [4] and a cepstro-temporal smoothing estimation algorithm of the *a priori* SNR [30]. The implementation of the algorithm is a MATLAB function provided by the author. Only one parameter, the minimum gain, is set as -20 dB for this function. This minimum gain value is the same as that in the proposed algorithm.

Fig. 10 shows the spectrogram of the processed noisy signal in Fig. 3. The color bars at the right side of the panels are in dB scale. Panel (a) shows the enhancement result by the ideal Wiener gain. This result is used as a reference for the results of the two algorithms. Panel (b) shows the result by applying the gain estimated from $\widehat{SNR}_v$ in (25). When applying this gain, only the periodic frames of the noisy speech are enhanced. Panel (c) shows the result by applying the gain calculated by (30). When applying this gain, both the periodic and the aperiodic frames of the noisy speech are enhanced. Panel (d) shows the result calculated by (31), which is the comb-filtered output of the result in panel (c). It can be seen that the result in (d) has reduced some noise between the harmonics in the periodic frames and shows a slightly clearer harmonic structure. Panel (e) shows the result calculated by the MMSE algorithm. As the MMSE algorithm uses the FFT transform to get the subbands, the noise levels between the harmonics are lower than that in panel (d). However, as the MMSE algorithm assumes that the noise level changes slower than speech, it cannot detect the level of the highly nonstationary noise like train noise. It can be seen that at the time around 0.45 s, the residual noise in (e) is stronger than that in (d).

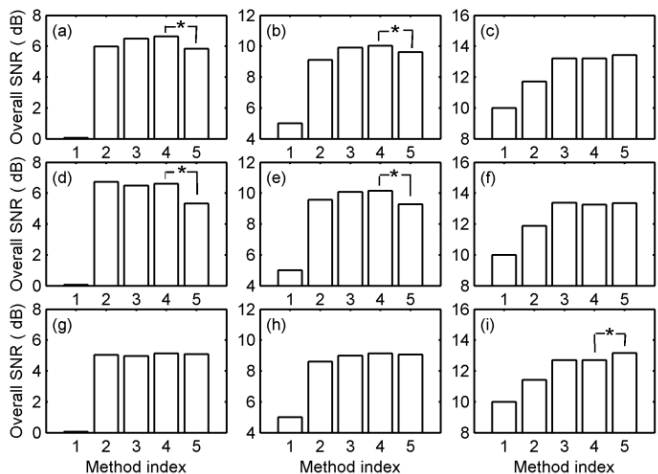

Fig. 11. Average overall SNR of original and processed noisy signals in car, train, and babble noise (rows) at overall SNR of 0, 5, and 10 dB (columns). The star denotes a significant difference (t-test, p <0.05) between method 4 and method 5. Method indexes 1, 2, 3, 4, and 5 correspond to the original noisy signal, the noisy signal resynthesized by applying the gain estimated from $\widehat{SNR}_v$ in (25), the noisy signal resynthesized by applying the gain in (30), the noisy signal resynthesized by (31), and the noisy signal processed by the MMSE method, respectively. (a, b, c) car noise at overall SNR of 0, 5, 10 dB.

(d, e, f) train noise at overall SNR of 0, 5, 10 dB. (g, h, i) babble noise at overall SNR of 0, 5, 10 dB.

Fig. 11 shows the average overall SNR of original and processed noisy signal in car, train, and babble noise at overall SNR of 0, 5, and 10 dB. The bars show the average values across the 30 sentences. Method indexes 1, 2, 3, 4, and 5 correspond to the original noisy signal, the noisy signal resynthesized by applying the gain estimated from $\widehat{SNR}_v$ in (25), the noisy signal resynthesized by applying the gain in (30), the noisy signal resynthesized by (31), and the noisy signal processed by the MMSE method, respectively. The processing delay for method 2, 3, and 4 is the sum of 16 ms introduced by gammatone filterbank and 32 ms introduced by frame-based processing, and the processing delay for method 5 is 32 ms introduced by frame-based processing. The panels (a), (b), and (c) show the signal in the relatively stationary car noise at overall SNR of 0, 5, and 10 dB, respectively; the panels (d), (e), and (f) show the signal in highly nonstationary train noise at overall SNR of 0, 5, and 10 dB, respectively; the panels (g), (h), and (i) show the signal in the babble noise at overall SNR of 0, 5, and 10 dB, respectively. The star denotes a significant difference (t-test, p < 0.05) between method 4 and method 5. Generally, method 2, which only enhances the periodic frames of the noisy speech, can achieve a higher average overall SNR compared to method 1 (unprocessed); method 3, which enhances both the periodic and aperiodic frames of the speech, can achieve a higher average overall SNR compared to method 2 (except in panel (d) and (g)); method 4, which applies a comb filtering processing to the output of method 3, can further slightly improve the average overall SNR at low SNR (0 dB and 5 dB); compared to the method 5 (MMSE algorithm), the t-test shows that method 4 gives significantly better improvement in car and train noise at the overall SNR of 0 and 5 dB (panel (a), (b), (d), and (e)), significantly less improvement in babble noise at overall SNR of 10 dB (panel (i)), and comparable (non-significantly different) improvement in the other cases. The improvement is less at 10 dB SNR, because in high SNR conditions the algorithm may reduce some speech components during voiced frames; similar results have also been found in other algorithms based on periodicity analysis (e.g., Fig. 19 in [7]).

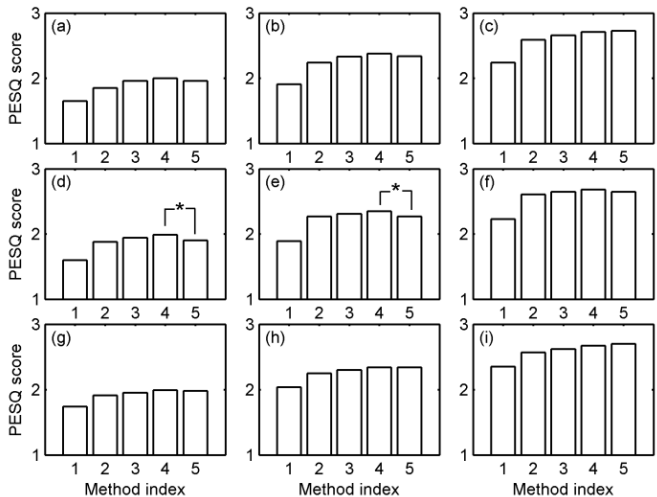

Fig. 12. Average PESQ score of original and processed noisy signal in car, train, and babble noise (rows) at overall SNR of 0, 5, and 10 dB (columns). The star denotes a significant difference (t-test, p<0.05) between method 4 and method 5. Method indexes and the panel labels represent the same conditions as in Fig. 11.

Fig. 12 shows the average PESQ score of original and processed noisy signal in car, train, and babble noise at overall SNR of 0, 5, and 10 dB. The bars show the average values across the 30 sentences. Method indexes and the panel labels represent the same conditions as in Fig. 11. The star denotes a significant difference (t-test, p < 0.05) between method 4 and method 5. Generally, method 2 (with an overall average PESQ score of 2.24; the overall average PESQ score means the score averages across all noise and SNR conditions) can achieve a higher average PESQ score compared to method 1 (with an overall average PESQ score of 1.96); method 3 (with an overall average PESQ score of 2.30) can achieve a higher average PESQ score compared to method 2; method 4 (with an overall average PESQ score of 2.35) can further improve the average PESQ score and achieves slightly higher score than the method 5 (with an overall average PESQ score of 2.32). Specifically, compared to the method 5, the t-test results show that method 4 gives significantly better improvement in train noise at the overall SNR of 0 and 5 dB (panel (d) and (e)), and comparable (non-significantly different) improvement in the other cases.

## IV. DISCUSSION AND CONCLUSION

The parameters in the proposed algorithm are mainly chosen empirically. Some of the parameters, including those in equation (13), (17), and (18), need further optimization in future studies. As it is hard to derive theoretical foundations for these parameters, this requires a large experimental study on related datasets, which goes beyond the scope of the current study.

The proposed online algorithm was compared to an algorithm that is representative for the class of statistical model-based algorithms, which also work in an online mode. It would be interesting to compare the proposed algorithm with other algorithms that use similar strategies of speech detection and separation as the proposed algorithm. However, as some of these algorithms may not be capable of blind [7] or online [31] processing, the comparison between these algorithms with the proposed algorithm may be biased. It would be more significant to derive online, blind-processing versions of these algorithms before comparing them with the proposed algorithm.

The algorithm proposed here divides the frames into periodic and aperiodic frames. This means that the algorithm performs a classification of voiced speech and unvoiced speech or noise, which can be seen as a voice activity detector (VAD). VADs based on speech periodicity features have been proposed earlier, e.g., [32]. When using VAD in the noise estimation for speech enhancement, the noise level is estimated by smoothing during frames without voice activity and kept constant during frames with voice activity. Different from this type of VAD, the proposed algorithm is able to estimate the noise level during frames with voice activity based on the relation between SNR and periodicity derived in this paper. This property seems important for an accurate estimation of non-stationary noise during voiced frames and was shown to achieve better speech enhancement according to the results presented here.

To detect and separate unvoiced speech components, a method similar to the approach taken in the statistical model based algorithm is adopted in the proposed algorithm. This method assumes that the noise changes slowly compared to speech; under this assumption, the noise level can be estimated with the recursive smoothing method, and the a priori SNR can be estimated by the simple decision-directed approach. This also means that during these aperiodic frames, any unit with sudden energy increase can be interpreted as unvoiced speech. To achieve better non-stationary noise suppression during unvoiced frames, some machine learning methods have been proposed. For example, Hu and Wang selected the features of unvoiced phonemes including spectral shape, intensity, and duration and classification algorithm to distinguish the unvoiced speech from background noise and achieved positive results [9]. Recently, some algorithms based on the deep neural network (DNN) framework achieved nearly perfect separation of speech from many types of non-stationary noise [33, 34]. Although the internal mapping function between noisy speech and clean speech in these algorithms is complicated, it would be interesting to analyze them and adopt their successful aspects into the knowledge-based algorithm proposed here.

Three representative types of noise are used here for evaluation. For the relatively stationary car noise, the state-of-the-art statistical-model based algorithms have achieved very good enhancement results and performed the best among current algorithms. So the proposed algorithm cannot be expected to provide further improvement for this condition. For the highly non-stationary train noise, however, the proposed algorithm outperforms the reference statistical-model based algorithm as expected. The proposed algorithm at present cannot deal with voiced components of the non-stationary babble noise and thus can only achieve comparable enhanced performance as the reference statistical-model based algorithm. An improvement of the proposed pitch detection and estimation algorithm to deal with multi-pitch conditions may help to improve the performance of the proposed algorithm in the babble noise.

In conclusion, this paper has introduced an online algorithm for frame-subband SNR estimation and speech enhancement based on the analysis of periodicity, estimation of noise level, and estimation of a *priori* SNR in speech. The algorithm achieves online-applicability by frame-by-frame signal analysis and processing. For each frame, the signal is decomposed into auditory frequency subbands by a novel IIR implementation of a phase-corrected complex-valued gammatone filterbank. The real-part of the filtered complex-valued output is the signal and the absolute value of the output is the Hilbert envelope of the signal. The subband signal can be summed up directly after the analysis and processing stages to form the enhanced signal. In the analysis stage, the novel combination of NAC and CFR is used as periodicity feature, named periodicity degree PD, for fundamental period detection and estimation, and subsequent SNR estimation. Based on the periodicity degree and using a specific tracking method, the fundamental period of the speech in aperiodic noise can be well detected. The theoretical relation between periodicity degree and SNR for each frame-subband unit was derived based on the uncorrelated assumption of the speech and the noise and the uncorrelated assumption of the noise and its delayed version. The calculated data fits the theoretical relation well and makes it possible to estimate SNR by periodicity degree. Based on the estimated SNR, the noise level during the periodic frames of the speech can be estimated. Combined with a recursive estimation of the noise level during the aperiodic frames of the speech, the continuous noise level was estimated. The a *priori* SNR is estimated based on the estimated noise level by a method similar to the classical directed-decision method. Based on the a *priori* SNR, a continuous gain was applied to the signal. The enhanced results show effective improvement in the objective criteria of overall SNR and PESQ score. Compared to a state-of-the-art statistical-model based algorithm, the proposed algorithm gives better evaluation results in the highly non-stationary train noise and comparable results in the relatively stationary car noise and non-stationary babble noise.

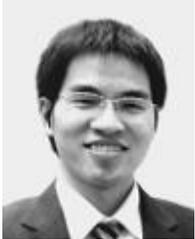

**Zhangli Chen** was born in Guangxi, China, in 1985. He received the B.Eng. and Ph.D. degrees in 2007 and 2012, both in biomedical engineering from Tsinghua University, Beijing, China.

He was a visiting Ph.D. student in the Hearing Group in University of Cambridge from 2010 to 2011. He is now a postdoctoral research scientist in the Department of Medical Physics and Acoustics, University of Oldenburg, Germany. His research interests include auditory modeling and speech signal processing.

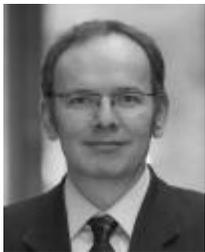

**Volker Hohmann** received the Physics degree (Dipl.-Phys.) and the doctorate degree in Physics (Dr. rer. nat.) from the University of Göttingen, Germany, in 1989 and 1993.

He has been a faculty member of the Physics Institute, University of Oldenburg, Germany since 1993 and was appointed full professor in 2012. His research expertise is in acoustics and digital signal processing with applications to signal processing in speech processing devices, e.g., hearing aids.

He is a consultant with the Hörzentrum Oldenburg GmbH. He was a Guest Researcher at Boston University, Boston, MA, (Prof. Dr. Colburn) in 2000 and at the Technical University of Catalonia, Barcelona, Spain in 2008. Prof. Hohmann received the Lothar-Cremer price of the German acoustical society (DEGA) in 2008 and the German President's Award for Technology and Innovation in 2012.